\documentclass[preprint,12pt]{article}
\usepackage{array,multirow,makecell}
\setcellgapes{1pt}
\makegapedcells
\newcolumntype{R}[1]{>{\raggedleft\arraybackslash }b{#1}}
\newcolumntype{L}[1]{>{\raggedright\arraybackslash }b{#1}}
\newcolumntype{C}[1]{>{\centering\arraybackslash }b{#1}}

\usepackage{amssymb}
\usepackage{hyperref}
\usepackage{mathtools}
\usepackage{xcolor}

\title{On the sound dispersion and attenuation in fluids due to thermal and viscous effects}

\author{Azeddine Zaidni, Saad Benjelloun \\ Université Mohammed VI Polytechnique}

\begin{document}

\maketitle

\begin{abstract}
In this paper, we derive a dispersion relation for sound waves in viscous and heat conducting fluids. In particular this dispersion (i.e. variation of speed of sound with frequency) is shown to be of second order of magnitude, w.r.t. Knudsen numbers, as in the Stokes \cite{Stokes} case, corresponding to non-conductive fluid (Prandtl number $Pr = \infty$). This formula completes the classical attenuation relation called Stokes-Kirchhoff.  We represent in a simplified manner the Kirchhoff approach to derive this attenuation \cite{Kirchhoff}, starting from the 3D compressible Navier-stokes system. The classical Stokes-Kirchhoff formula has been questioned recently in \cite{Hu} and a different (and incorrect) formula was proposed. We point out the non-trivial assumptions that are violated in the new derivation in \cite{Hu} to reestablish the classical Stokes-Kirchhoff formula. Finally, we give an explanation to differences in dispersion and attenuation formulae that one may find in the literature through analysing the form of the considered attenuated solutions.
\end{abstract}






\tableofcontents

\section{Introduction}\label{sec:intro}
A first study of sound wave propagation in viscous fluids was published by G. Stokes in 1845 \cite{Stokes}, and quantifies the attenuation of the amplitude of sound wave as well as its dispersion (i.e. the variation of the speed of sound with the frequency). Stokes' law applies in an isotropic and homogeneous medium without taking into account heat conductivity $Pr = \infty$. A generalization of this study taking into account the thermal conductivity was proposed by G.Kirchhoff in 1868 \cite{Kirchhoff}. The Stokes-Kirchhoff relation expresses the sound attenuation in function of the characteristics of the fluid, namely the density $\rho$, the dynamic viscosity $\mu$, the thermal conductivity $\nu$, and the sound wave frequency $\omega$. 


The movement of real fluids is governed by the 3D compressible Navier-stockes system with viscosity and thermal conduction terms :
\begin{equation}
\left\{\begin{aligned}
\frac{\partial \rho}{\partial t}+\frac{\partial (\rho u_{j})}{\partial x_{j}} &=0 \\
\frac{\partial (\rho u_{i})}{\partial t}+\frac{\partial (\rho u_{i} u_{j} )}{\partial x_{j}}+\frac{\partial p}{\partial x_{i}} &=\frac{\partial}{\partial x_{j}} S_{i j} \\
\frac{\partial (\rho E)}{\partial t}+\frac{\partial (\rho E u_{j}) }{\partial x_{j}}+\frac{\partial (p u_{j})}{\partial x_{j}} &=\frac{\partial}{\partial x_{j}}\left( S_{i j} u_{i}\right)-\frac{\partial q_{j}}{\partial x_{j}}
\end{aligned}\right.
\label{NS3D}
\end{equation}
where $\Vec{u}=(u_1,u_2,u_3)$ is the velocity vector and $S$ is the deviator part of the strain rate tensor:
$$S_{i j}=\mu\left(u_{i, j}+u_{j, i}\right)+\mu^{\prime} u_{k, k} \delta_{i j}\,\,,\,\,\,\,\,\, u_{i, j}=\frac{\partial u_{i}}{\partial x_{j}}$$
where we denote $\mu$ the dynamic viscosity, and
$\mu^{\prime}$ the second viscosity. The vector $q$ is the heat flux, calculated using Fourier's law, $q=-\lambda \nabla T$, where $\lambda$ is the thermal conductivity, and $T$ the temperature. $E$ is the total energy $e + \frac{||\mathbf{u}||^2}{2}$, with $e$ the fluid specific internal energy.

The source term in the momentum equation can be wtitten
$$
\frac{\partial}{\partial x_{j}} S_{i j} =\mu^{} \Delta u_i + (\mu + \mu^{\prime})\frac{\partial}{\partial x_i}\operatorname{div}(\Vec{u}) 
$$

and the system \eqref{NS3D} becomes:

$$
\left\{\begin{aligned}\frac{\partial{\rho}}{\partial t} + \rho \operatorname{div}(\Vec{u}) + \Vec{u}.\Vec{\operatorname{grad}}(\rho) &= 0\\
u_i \frac{\partial \rho}{\partial t} + \rho\frac{\partial u_i}{\partial t} + u_i \operatorname{div}(\rho\Vec{u})+\rho u_j\frac{\partial u_i}{\partial x_j} + \frac{\partial p}{\partial x_i}&= \mu^{} \Delta u_i + (\mu + \mu^{\prime})\frac{\partial}{\partial x_i}\operatorname{div}(\Vec{u})\\
\rho\frac{\partial e}{\partial t}+\rho u_j\frac{\partial e}{x_j} + u_j\frac{\partial p}{\partial x_j} + p\operatorname{div}(\Vec{u}) + \rho \Vec{u}.\frac{\mathrm{D}\Vec{u}}{\mathrm{D}t} &=\frac{\partial}{\partial x_{j}}\left(  S_{i j} u_{i}\right)-\frac{\partial q_{j}}{\partial x_{j}}\end{aligned}\right.
$$
where  $\frac{\mathrm{D}\Vec{u}}{\mathrm{D}t}$ is the particular derivative.

Using the thermodynamic identity  \eqref{A0} $d e=T d s+\frac{p}{\rho^{2}} d \rho$ \cite{thermo} we can replace the energy equation by an equation for the entropy $s$:

\begin{equation}
\left\{\begin{aligned}\frac{\partial{\rho}}{\partial t} + \rho \operatorname{div}(\Vec{u}) + \Vec{u}.\Vec{\operatorname{grad}}(\rho) &= 0\\
 \rho\frac{\partial u_i}{\partial t} +\rho u_j\frac{\partial u_i}{\partial x_j} + \frac{\partial p}{\partial x_i}&= \mu^{} \Delta u_i + (\mu + \mu^{\prime})\frac{\partial}{\partial x_i}\operatorname{div}(\Vec{u})\\
\rho T \frac{\partial s}{\partial t}+\rho T u_j\frac{\partial s}{x_j} + u_j\frac{\partial p}{\partial x_j} + \rho \Vec{u}.\frac{\mathrm{D}\Vec{u}}{\mathrm{D}t} &=u_i\frac{\partial S_{i j}}{\partial x_{j}}  +  S_{i j} \frac{\partial u_i}{\partial x_{j}}-\frac{\partial q_{j}}{\partial x_{j}}\end{aligned}\right.
\label{NS3D_bis}
\end{equation}
If we multiply the second equation by $u_i$ and obtain:
$$u_i\frac{\partial p}{\partial x_i} + \rho \Vec{u}.\frac{\mathrm{D}\Vec{u}}{\mathrm{D}t} = \mu^{} u_i \Delta u_i + (\mu + \mu^{\prime})u_i \frac{\partial}{\partial x_i}\operatorname{div}(\Vec{u}) = u_i\frac{\partial S_{i j}}{\partial x_j} $$                        
Hence :
\begin{equation}
\left\{\begin{aligned}\frac{\partial{\rho}}{\partial t} + \rho \operatorname{div}(\Vec{u}) + \Vec{u}.\Vec{\operatorname{grad}}(\rho) &= 0\\
 \rho\frac{\partial u_i}{\partial t} +\rho u_j\frac{\partial u_i}{\partial x_j} + \frac{\partial p}{\partial x_i}&= \mu^{} \Delta u_i + (\mu + \mu^{\prime})\frac{\partial}{\partial x_i}\operatorname{div}(\Vec{u})\\
\rho T \frac{\partial s}{\partial t}+\rho T u_j\frac{\partial s}{x_j}  &=  S_{i j} \frac{\partial u_i}{\partial x_{j}}-\frac{\partial q_{j}}{\partial x_{j}}\end{aligned}\right.
\label{NS3D_bis1}
\end{equation}

In the next section, we derive the Stokes-Kirchhoff attenuation rate from the linearized version of the Navier-stokes system \eqref{NS3D_bis1}. We also derive the sound dispersion, i.e. the variation of the speed of propagation of the sound wave with the frequency. In section 3, we will give a presentation of Kirchhoff derivation as this is to our knowledge not available in the literature, and as the original paper of Kirchhoff is only available in German. Finally in section 4 we will present the derivation method by Hu \cite{Hu}, to make clearer the issue in this derivation, in a more complete way than has been done in \cite{Jordan} and by exhibiting the role played by Knudsen dimensionless numbers.


\section{Derivation of sound wave dispersion and attenuation}

In this section we will derive the Stokes-Kirchhoff relation directly from the Navier-Stokes system, using a convenient matrix formulation. We also obtain the dispersion relation giving the sound wave propagation speed as function of the frequency.

The linearization of the system \eqref{NS3D_bis1} around a constant solution  $(\rho_0 , \Vec{u}=\Vec{0}, T_0)$ reads:

$$
\left\{\begin{aligned}\frac{\partial{\rho}}{\partial t} + \rho_0 \operatorname{div}(\Vec{u}) &= 0\\
\rho_0\frac{\partial u_i}{\partial t} +\frac{\partial p}{\partial x_i}&= \mu^{} \Delta u_i +(\mu+\mu^{\prime}) \frac{\partial}{\partial x_i}\operatorname{div}(\Vec{u})\\
\rho_0 T_0\frac{\partial s}{\partial t}  &= \lambda \Delta T\end{aligned}\right.
$$

Using  $dp = \rho_0 \Gamma C_v dT + c_T^2 d\rho$, and $d T=\frac{\Gamma T}{\rho} d \rho+\frac{T}{C_{v}} d s$, \cite{thermo} with $\Gamma$ the Gr\"uneisen  coefficient \cite{thermo}, $C_v$ the  isochoric heat capacity, and $c_T$ the isothermal speed of sound, all at the state $(\rho_0, T_0)$, the linearized system becomes:
$$\left\{\begin{aligned}
\frac{\partial{\rho}}{\partial t} + \rho_0 \operatorname{div}(\Vec{u}) &= 0\\ \frac{\partial u_i}{\partial t} +\Gamma C_v \frac{ \partial T }{ \partial x_i }  + \frac{c_T^2}{\rho_0} \frac{\partial \rho }{\partial x_i} &=(\mu^{}/\rho_0) \Delta u_i +\frac{(\mu+\mu^{\prime})}{\rho_0} \frac{\partial}{\partial x_i}\operatorname{div}(\Vec{u})\,\,\,\, , i=1,2,3\\
\frac{\partial T}{\partial t} +  \Gamma T_0 \operatorname{div}(\Vec{u}) &=   \frac{\lambda}{\rho_0 \,C_v}  \Delta T
\end{aligned}\right.$$

This linear differential system  is of the form:
$$
\frac{\partial W}{\partial t}+A \frac{\partial W}{\partial x_1}+B \frac{\partial W}{\partial x_2}+C \frac{\partial W}{\partial x_3}=D \Delta{ W} + \sum_{1 \leq i,j\leq 3 }E_{i,j} \frac{\partial^2 W}{\partial x_i \partial x_j} 
$$
where $W = {}^{t}(\rho, u_1, u_2, u_3, T )$, and $(A,B,C,D,E_{i,j})$ are $5 \times 5$ matrices.\\

We consider here plane waves and we suppose for example that the wave propagates in the $x_1$ direction. The solution wave does not depend on the $x_2$ and $x_3$ coordinates, and $u_2 = u_3 = 0$:

$${ W(x_1,x_2,x_3,t)=W(x_1,t)}$$
We get:
$$
\frac{\partial W}{\partial t}+A \frac{\partial W}{\partial x_1} = \left(D  + E_{1,1}\right) \frac{\partial^2 W}{\partial x_1^2 } 
$$
The matrices $(A,D,E_{1,1})$ are given by:
$$
A=\left(\begin{array}{ccccc}
0 & \rho_{0} & 0 & 0 & 0\\
\frac{c_{T}^{2}}{\rho_{0}} & 0 & 0 & 0 &\Gamma C_{v} \\
0 & 0 & 0 & 0 & 0\\
0 & 0 & 0 & 0 & 0\\
0 & \Gamma T_0 & 0 & 0 & 0
\end{array}\right), \quad D=\left(\begin{array}{ccccc}
0 & 0 & 0 & 0 & 0\\
0 & \frac{\mu^{\prime}}{\rho_0} & 0 & 0 & 0\\
0 & 0 & 0  & 0 & 0\\
0 & 0 & 0 & 0  & 0\\
0 & 0 & 0 & 0 & \frac{\lambda}{\rho_0 C_v}
\end{array}\right) 
$$
$$
E_{1,1}=\left(\begin{array}{ccccc}
0 & 0 & 0 & 0 & 0 \\
0 & \frac{(\mu + \mu^{\prime})}{\rho_0} & 0 & 0 & 0\\
0 & 0 & 0 & 0 & 0\\
0 & 0 & 0 & 0 & 0\\
0 & 0 & 0 & 0 & 0
\end{array}\right)$$

We look for non-zero harmonic plane-wave solutions of the form:
\begin{equation}\label{sol1}
W=W_{0} \exp (i\omega t + k x) = W_{0} \exp ( k_R x) \exp \left(i (\omega t + k_I x) \right),
\end{equation}

where $\omega$ is the sound frequency and $k=k_R + i k_I$ is a complex wavelength, that contains the wavelength $k_I$ and the attenuation $k_R$. In section $\ref{sec:disc}$ we will discuss this choice for the form of the solution.

We get then: 
$$\left[ \omega I- ik A+ik^{2} (D+E_{1,1})\right]W=0$$

Then non-zero solutions exists if and only if :
$$
\mathbf{det}(\omega I- ik A+ik^{2} (D+E_{1,1})) = 0,
$$
which gives the bi-quadratic polynomial on $k$:
{\small
$$
\omega^2 \left(\omega^{3}+k^2 \left[ \frac{i }{\rho}\left({(2\mu + \mu^{\prime}}+\frac{\lambda}{C_{v}}\right){\omega}^{2} + c^2\omega\right]  +
k^4\left[\frac{i \lambda}{\rho}\left(\frac{c^{2}}{C_{v}}-\Gamma^{2} T\right) - \frac{(2\mu + \mu^{\prime})  \lambda }{ \rho^{2} C_{v}}\omega \right]\right)=0$$
}

As we have \eqref{A4}\,\,\,  $\Gamma^{2} C_{v} T=\frac{\gamma-1}{\gamma} c^{2} = c^2 -  c_T^2$, we can write the dispersion relation as :
{\small
\begin{equation}\label{dispersion}
\omega^{3}+k^2 \left[ \frac{i }{\rho}\left({(2\mu + \mu^{\prime}}+\frac{\lambda}{C_{v}}\right){\omega}^{2} + c^2\omega\right]  +
k^4\left[\frac{i \lambda c^2}{\rho C_v \gamma} - \frac{(2\mu + \mu^{\prime})  \lambda }{ \rho^{2} C_{v}}\omega \right]=0
\end{equation}
}
which we can also put in the dimensionless form 
{\small
\begin{equation}\label{dispersion_AD}
1 + \left(\frac{kc}{\omega}\right)^2 \left[ \frac{i }{\rho}\left({(2\mu + \mu^{\prime}}+\frac{\lambda}{C_{v}}\right) \frac{\omega}{c^2} + 1\right]  +
\left(\frac{k c}{\omega} \right)^4 \left[\frac{i \lambda \omega}{\rho  c^2 C_v \gamma} - \frac{(2\mu + \mu^{\prime})  \lambda \omega^2 }{ \rho^{2} C_{v}  c^4 } \right]=0
\end{equation}
}

Introducing the Knudsen numbers $K_n=\frac{(2\mu + \mu')\omega}{\rho c^2}$ and $K_{T} = \frac{\lambda \omega}{\rho c^2 C_p}$ as in \cite{SBJMG}, we have :
{\small
\begin{equation}\label{dispersion_ADKn}
1 + \left(\frac{kc}{\omega}\right)^2 \left[ i \left( K_n + \gamma K_{T}\right) + 1\right]  + \left(\frac{k c}{\omega} \right)^4 \left[i K_{T} - \gamma\, K_n \, K_{T} \right]=0
\end{equation}
}

We note that as per the continuum hypothesis, the Knudsen numbers $K_n$ and $K_{T}$ are supposed very small ($\approx 10^{-2}$).

\subsection{Attenuation and dispersion }

In this session we will express the four roots $( k_1^{\pm},k_2^{\pm})$ of the bi-quadratic polynomial \eqref{dispersion}, and give a Taylor expansion, at order $2$ in $\epsilon = K_n + K_{th} = \frac{\omega \left(2\mu +\mu^{\prime} + \frac{\lambda}{C_p} \right)}{\rho c^2} \ll 1$. The polynomial discriminant is given by:  
$$ D = \frac{c^4}{\omega^4} \left( 1 + \frac{2\,i\omega(\frac{2\mu+\mu^{\prime}}{{ \rho}}+\frac{\lambda}{\rho\,C_v})}{c^2} -   4\frac{i\omega \lambda}{\rho C_v \gamma \, c^2 }  - \frac{\omega^2 (\frac{2\mu + \mu^{\prime}}{\rho}-\frac{\lambda}{\rho C_v})^2}{c^4}   \right)$$
Using the expansion $\sqrt{1+\epsilon} =  1 + \frac{\epsilon}{2} -\frac{\epsilon^2}{8} + O(\epsilon^3)$ we get: \\
$\begin{aligned}
\sqrt{ D }= \frac{c^2}{\omega^2}& \left( 1 + \frac{i\omega(\frac{2\mu+\mu^{\prime}}{{\rho}}+\frac{\lambda}{\rho C_v})}{c^2}- 2\frac{i\omega \lambda}{\rho C_v\gamma \, c^2 } \right. \\ & \left. + \frac{2 \omega^2(2\mu+\mu^{\prime}) \lambda}{\rho^2 C_v c^4}  + \frac{2 \omega^2 \lambda^2}{\rho^2 C_v^2 \gamma^2 \, c^4 }  - \frac{ 2 \omega^2 \lambda(\frac{2\mu+\mu^{\prime}}{\rho}+\frac{\lambda}{\rho C_v})}{\rho C_v\gamma c^4}   + O(\varepsilon^3) \right) \end{aligned}$\\
Then

$$ 
(k_1^{\pm})^2 =  -\frac{c^2}{2\omega^2} \left( 1 + \frac{i\omega(\frac{2\mu+\mu^{\prime}}{\rho}+\frac{\lambda}{\rho C_v})}{c^2}\right) - \frac{\sqrt{D}}{2}.
$$

We obtain then the order $2$ expansion :

$$\begin{aligned}
k_1^{\pm} = \pm \frac{i\omega}{c}&\left(1
-
\frac{i\omega(\frac{2\mu+\mu^{\prime}}{\rho}+\frac{\lambda}{\rho C_v})}{2c^2 }
+
\frac{i\omega\lambda}{2\rho C_v\gamma c^2} 
- 
\frac{3\omega^2(\frac{(2\mu+\mu^{\prime})^2}{\rho^2}+(\frac{\lambda}{\rho C_v})^2)}{8c^4}\right.\\
&-\left.
\frac{\omega^2 (2\mu+\mu^{\prime})\lambda}{4\rho^2 C_v c^4}
-
\frac{7\omega^2\lambda^2}{8\rho^2 C_v^2 \gamma^2 c^4} 
+
\frac{\omega^2\lambda(\frac{2\mu+\mu^{\prime}}{\rho}+\frac{\lambda}{\rho C_v})}{2\rho C_v\gamma c^4}  
+ O(\varepsilon^3)\right)
\end{aligned}$$
On the other side, for the other pair of solutions: 
$$(k_2^{\pm})^2= -\frac{c^2}{2\omega^2} \left( 1 + \frac{i\omega(\frac{2\mu+\mu^{\prime}}{\rho}+\frac{\lambda}{\rho C_v})}{c^2}\right) + \frac{\sqrt{D}}{2}$$
Hence, at order $2$ in $\varepsilon$:
$$(k_2^{\pm})^2=  \frac{c^2}{2\omega^2} \left(- \frac{i\omega \lambda}{\rho C_v\gamma \, c^2 }  + \frac{ \omega^2(2\mu+\mu^{\prime}) \lambda}{\rho^2 C_v c^4}  + \frac{ \omega^2 \lambda^2}{\rho^2 C_v^2 \gamma^2 \, c^4 }  - \frac{  \omega^2 \lambda(\frac{2\mu+\mu^{\prime}}{\rho}+\frac{\lambda}{\rho C_v})}{\rho C_v\gamma c^4}   + O(\varepsilon^3) \right)$$
$$ = \frac{c^2}{2\omega^2}\left(-iK_{th} + (\gamma-1)K_{th}K_n -(\gamma -1)K_{th}^2 + O(\varepsilon^3)\right)$$

Hence,
$$ k_2^{\pm} = \pm (1+i)\frac{c\,\sqrt{K_{th}}}{\sqrt{2}\,\omega} \left(1+\frac{i(\gamma-1)(K_n + K_{th})}{2} + \frac{(\gamma-1)^2 (K_n + K_{th})^2}{8} + O(\varepsilon^2) \right)$$


Only the first pair of solutions $k_1$ corresponds to sound waves propagation and the attenuation $k_R$  in the $x>0$ direction is given by:
$$k_R = Re(k_1^{-}) = -\frac{\omega^{2}}{2 c^{3}}\left[\frac{2\mu+\mu^{\prime}}{\rho}+\frac{\lambda}{\rho C_v}\left(1-\frac{1}{\gamma}\right)\right]$$

The above relation is exactly the Stokes-Kirchhoff attenuation derived in \cite{Kirchhoff}. We note that the attenuation in the medium of propagation varies with the frequency.
Concerning sound wave dispersion, we calculate the phase velocity $\frac{\omega}{|Im(k_1^{\pm})|}$ 
$$
v_s(\omega) = \frac{\omega}{|Im(k_1)|} = \frac{c}{1
  - 
\frac{3\omega^2(\frac{(2\mu+\mu^{\prime})^2}{\rho^2}+\frac{\lambda^2}{\rho^2 C_v^2})}{8c^4}
-
\frac{\omega^2 (2\mu+\mu^{\prime})\lambda}{4\rho^2 C_v c^4}
-
\frac{7\omega^2\lambda^2}{8 \rho^2 C_v^2 \gamma^2 c^4} 
+
\frac{\omega^2\lambda((\frac{2\mu+\mu^{\prime}}{\rho})+\frac{\lambda}{\rho C_v})}{2\rho C_v \gamma c^4}  }
$$

{$$
= \frac{c}{1
  - 
\frac{3}{8}{K_n}^2
-
\left(\frac{3}{8}\gamma^2 +\frac{7}{8}-\frac{\gamma}{2}\right)K_{th}^2
-\left(\frac{\gamma}{4}-\frac{1}{2}\right)K_n K_{th}  }
$$}\\

We see that the dispersion is of order $2$ in $\epsilon$.

In the case where the thermal conductivity is zero $(\lambda=0)$ we have $$v_s(\omega) = \frac{\omega}{|Im(k_1^{-})|} = {\frac{c}{1 - \frac{3}{8}K_n^2} = } \frac{c}{1 -\frac{3\omega^2(2\mu+\mu^{\prime})^2}{8\rho^2c^4}} \approx c\left(1 +  \frac{3\omega^2(2\mu+\mu^{\prime})^2}{8\rho^2 c^4}\right)$$

This formula is different form the dispersion rate derived by Stokes in \cite{Stokes}. This is because Stokes uses a different form of solutions as we will explain in section \ref{sec:disc}.

\section{Review of Kirchhoff derivation.}
\subsection{Kirchhoff dimensionless system}

In this section we present the derivation of the Stokes-Kirchhoff attenuation formula according to the classical Kirchhoff paper \cite{Kirchhoff}. This derivation addresses first order terms only and hence does not show the sound waves dispersion. A consequent part of Kirchhoff derivation in \cite{Kirchhoff} aims to establish the energy equation from first thermodynamical principles which makes the paper hard to read. We represent here this derivation more simply, starting from the linearized 3D compressible Navier-stockes system, with viscosity and thermal conduction terms, as obtained in section 1:

$$\left\{\begin{aligned}
\frac{\partial{\rho}}{\partial t} + \rho_0 \operatorname{div}(\Vec{u}) &= 0\\ \rho_0\frac{\partial u_i}{\partial t} +\rho_0\Gamma C_v \frac{ \partial T }{ \partial x_i }  + c_T^2 \frac{\partial \rho }{\partial x_i} &=\mu^{} \Delta u_i +(\mu+\mu^{\prime}) \frac{\partial}{\partial x_i}\operatorname{div}(\Vec{u})\,\,\,\, , i=1,2,3\\
\frac{\partial T}{\partial t} +  \Gamma T_0 \operatorname{div}(\Vec{u}) &=   \frac{\lambda}{\rho_0 \,C_v}  \Delta T
\end{aligned}\right.$$

Kirchhoff \cite{Kirchhoff} introduces the notations below :

$$
\mu_1 = \frac{\mu}{\rho}, \,
\mu_2 =\frac{\mu + \mu^{\prime}}{\rho}, \,
\sigma = \frac{\rho}{\rho_0},\,
\nu = \frac{\lambda}{\rho C_v},\,
\theta = \frac{\alpha}{\gamma-1} T = \frac{\alpha \, C_v}{C_p-C_v} T = \frac{\alpha \,  c_T^2}{c^2 -c_T^2} T,$$
with $\alpha$ the isobaric expansive coefficient (in $K^{-1}$) verifying \eqref{A4}
$$
\alpha =  \left.\frac{1}{v}\frac{\partial v}{\partial T} \right)_p = -\left. \frac{1}{\rho}\frac{\partial \rho}{\partial T} \right)_p = \frac{\Gamma C_v}{ c_T^2 } = \frac{1}{\Gamma T_0} \frac{c^2 - c_T^2}{ c_T^2 }.
$$

Hence, $\theta = \frac{\Gamma C_v}{c^2 - c_T^2} T = \frac{T}{\Gamma\, T_0} $ and the linear Navier-Stokes system writes in these variables as :
\begin{eqnarray}
&&\frac{\partial \sigma}{\partial t}+\operatorname{div}(\Vec{u})=0 \\ 
&&\frac{\partial u_i}{\partial t}+c_T^{2} \frac{\partial \sigma}{\partial x_i}+\left(c^{2}-c_T^{2}\right) \frac{\partial \theta}{\partial x_i}=\mu_1\Delta u_i + \mu_2\frac{\partial}{\partial x_i}\operatorname{div}(\Vec{u})  \\
&&\frac{\partial \theta}{\partial t} + \operatorname{div}(\Vec{u}) = \nu  \Delta \theta
\end{eqnarray}
Or equivalently,
\begin{eqnarray}
&&\frac{\partial \sigma}{\partial t}+\operatorname{div}(\Vec{u})=0 \nonumber \\ 
&&\frac{\partial u_i}{\partial t}+c_T^{2} \frac{\partial \sigma}{\partial x_i}+\left(c^{2}-c_T^{2}\right) \frac{\partial \theta}{\partial x_i}=\mu_1\Delta u_i - \mu_2  \frac{\partial^{2} \sigma}{\partial x_i \partial t  }    \label{syskirchhoff} \\
&&\frac{\partial \theta}{\partial t} -\frac{\partial \sigma}{\partial t} = \nu  \Delta \theta \nonumber
\end{eqnarray}

\subsection{Kirchhoff derivation for attenuation rate}
Kirchhoff's method also consists in looking for solutions of type $Y(x,t) = Y(x) \exp{(h t)} = Y(x) \exp{(i\omega t)} $. Kirchhoff transforms first the linearized system (\ref{syskirchhoff}) into a single scalar bi-Laplacian equation on the variable $\theta$. Then one can look for solutions of the particular form in $x$  $$ \theta(t,x) = \theta(x) \exp{(i\,\omega t)} = \Theta \exp{(k x)}\, \exp{(i\,\omega t)}.$$

The relation of $k(\omega)$ gives the attenuation $Re(k)$, and the dispersive sound velocity is $v_s(\omega) = \frac{\omega}{ |Im(k)|}$.

From the system (\ref{syskirchhoff}), the spacial evolution of the variables is governed by the system :

\begin{eqnarray}
\operatorname{div}(\Vec{u})+h\, \sigma &=& 0 \label{u_sigma_1} \\
h u_i-\mu_1 \Delta u_i &=&-\frac{\partial Q}{\partial x_i} \label{u_Q_1}  \\
\sigma & = &\theta-\frac{\nu}{h} \Delta \theta \label{sigma_theta_1},
\end{eqnarray}

where $ Q=\left(c_T^2 + h \mu_2\right)\sigma
+\left(c^{2}-c_T^{2}\right) \theta $. Using \eqref{sigma_theta_1} we get: 
$$
Q =\left(c^2+h \mu_2\right)\theta
-\left(c_T^{2}+h \mu_2\right) \frac{\nu}{h} \Delta \theta.
$$

From \eqref{u_sigma_1} and \eqref{sigma_theta_1} we obtain:
\begin{equation}
\operatorname{div}(\Vec{u}) = - h\, \sigma = - h\theta + \nu \Delta \theta
\label{ux_theta_1}
\end{equation}

On other hand, we derive \eqref{u_Q_1} with respect to $x_i$  and we sum over $i$ to obtain:
$$
h \operatorname{div}(\Vec{u}) -\mu_1 \Delta (\operatorname{div}(\Vec{u})) = - \Delta Q \label{dxu_Q_1}.
$$
Finally, replacing  $\operatorname{div}(\Vec{u}) $ by the expression \eqref{ux_theta_1}, we get: 
$$
h^{2} \theta-\left[c^{2}+h\left(\mu_1+\mu_2+\nu\right)\right] \Delta \theta+\frac{\nu}{h}\left[c_T^{2}+h\left(\mu_1+\mu_2\right)\right] \Delta \Delta \theta=0
$$

We look for solutions of type $\theta(t,x) = \Theta \exp{(k x)} \exp{(ht)} $, and the characteristic polynomial for the scalar bi-laplacian equation is:
\begin{equation}
h^{2}-\left[c^{2}+h\left(\mu_1+\mu_2+\nu\right)\right]k^2
+\frac{\nu}{h}\left[c_T^{2}+ h \left(\mu_1+\mu_2\right)\right] k^{4}=0, \label{polyLambda}
\end{equation}
which is exactly the same as \eqref{dispersion_AD} and \eqref{dispersion_ADKn}, and that Kirchhoff writes in the form:
\begin{equation}
\frac{1}{k^4}-
\frac{\left[c^{2}+h\left(\mu_1+\mu_2+\nu\right)\right]}{h^{2}} \frac{1}{k}
+\frac{\nu}{h^3}\left[c_T^{2}+ h \left(\mu_1+\mu_2\right)\right] =0 \label{polyLambda-1}.
\end{equation}

If we take $\lambda = k^2$ and note $\mu = \mu_1 + \mu_2$, the roots of the polynomial $\lambda_{1}$ and
$\lambda_2$ verify:
\begin{eqnarray}
\frac{1}{\lambda_{1}}+\frac{1}{\lambda_{2}}
= \frac{c^2}{h^2} \left(  1 + \frac{h(\mu+\nu)}{c^2}\right)
\\
\frac{1}{\lambda_{1}} \frac{1}{\lambda_{2}}= \frac{c^2}{h^2} \left( \frac{\nu}{\gamma \,h} + \frac{\mu \,\nu}{c^2} \right)
\end{eqnarray}

We look for the approximate values of the roots, by an iterative process. If we suppose that $\mu_1$, $\mu_2$ and $\nu$ are of the same order, or more precisely if $\epsilon=\frac{h( \mu + \nu)}{c^2} = K_n + \gamma K_T \ll 1$, then we can suppose that one solution, say $\lambda_2$ is very large (order $\frac{h^2}{c^2\,\epsilon}$) with respect to $\lambda_1$ (order $\frac{h^2}{c^2}$). The first equation becomes at order 0:
$$\frac{1}{\lambda_1} = \frac{c^2}{h^2} + O(\epsilon).$$
and the second gives at order $1$:
$$\frac{1}{\lambda_2} = \lambda_1\,\frac{\nu\,c_T^2}{h^3} = \frac{\nu\,c_T^2}{h\,c^2} = \frac{\nu}{\gamma \,h} = \frac{c^2}{h^2}  \frac{1}{\gamma}\frac{h\,\nu}{\,c^2} + O(\epsilon^2).$$

Coming back to $\lambda_1$ we find:
$$\frac{1}{\lambda_{1}}=\frac{c^{2}}{h^{2}}+\frac{\mu+\nu}{h}-\frac{1}{\lambda_{2}} = \frac{c^{2}}{h^{2}}+\frac{\mu+\nu}{h}-\frac{\nu\,c_T^2}{h\,c^2} =  \frac{c^{2}}{h^{2}}+\frac{\mu}{h}+\frac{\nu}{h}(1 - \frac{c_T^2}{c^2}),$$
then
$$\frac{1}{\lambda_1} = \frac{c^2}{h^2}\left[ 1 + \frac{h}{c^2}\left(\mu + \nu \left(1-\frac{1}{\gamma}\right)\right)\right].$$

Hence we obtain complex wavelength
\begin{equation}
k_1^{\pm} = \pm \sqrt{ \lambda_{1}}=\pm\left\{\frac{h}{c}-\frac{h^{2}}{2 c^{3}}\left[\mu+\nu\left(1-\frac{1}{\gamma}\right)\right]\right\}\\
=\pm\left\{\frac{i\omega}{c}+\frac{\omega^{2}}{2 c^{3}}\left[\mu+\nu\left(1-\frac{1}{\gamma}\right)\right]\right\}
\end{equation}
corresponding the to classical Stokes-Kirchhoff attenuation formula. We point out that contrarily to the main statement in \cite{Hu} the derivation above due to Kirchhoff \cite{Kirchhoff} is strictly rigorous and correct and the iterative process to obtain the first order expansions does not include any error.

\section{Review of Hu et al \cite{Hu} derivation}

In this section we represent a derivation of a 'modified' Stokes-Kirchhoff formula according to the paper of Hu \cite{Hu}.
Starting from the linearized 1D compressible Navier-stockes system, on can derive the fully thermally-mechanically coupled equation set used by Hu \cite{Hu}. We present the derivation in the next subsection. 
\subsection{Thermally-mechanically coupled equation set}
Let the linearized 1D compressible Navier-stockes system:
\begin{equation} 
 \frac{\partial u}{\partial x}= - \frac{1}{\rho_0} \frac{\partial \rho}{\partial t}
 \label{NSL1c}
\end{equation}

\begin{equation}
\frac{{\partial} u}{\partial t}+ \frac{1}{\rho_0} \frac{\partial {p}}{\partial x}=\mu  \frac{\partial^{2} u}{\partial^{2} x}\label{NSL2c}
\end{equation}

\begin{equation}
\frac{\partial T}{\partial t}+  \Gamma T_0 \frac{\partial u}{\partial x} = 
 \frac{\partial}{\partial x}\left(\nu\,\frac{\partial T}{\partial x}\right)
\label{NSL3c}
\end{equation}
From the equation (\ref{NSL2c}) derived with respect to $x$ and (\ref{NSL3c}) we obtain:
$$\begin{cases}
\frac{\partial}{\partial t}(\frac{\partial u}{\partial x}) + \frac{1}{\rho_{0} } \frac{\partial^2 p}{\partial x^2}=\mu \frac{\partial^{3} u}{\partial x^3} \label{NSL2cx}\\
\frac{\partial T}{\partial t} +  \Gamma T_0 \frac{\partial u}{\partial x} = 
\nu  \left(\frac{\partial^2 T}{\partial x^2}\right)
\end{cases}
$$

We replace  $\frac{\partial u}{\partial x} $ by  $ -\frac{1}{\rho_0} \frac{\partial \rho}{\partial t}$ to get:
$$
\begin{cases}
- \frac{1}{\rho_0} \frac{\partial^2 \rho}{\partial t^2} + \frac{1}{\rho_0}\frac{\partial^2 {p}}{\partial x^2} = - \frac{\mu}{\rho_0} \frac{\partial}{\partial t} \left(\frac{\partial^2 \rho}{\partial x^2}\right)\\
\frac{\partial T}{\partial t} - \frac{\Gamma T_0}{\rho_0} \frac{\partial \rho}{\partial t} =
\nu \left(\frac{\partial^2 T}{\partial x^2}\right)
\end{cases}
$$

Using \eqref{A2}
$$ d \rho  = \frac{1}{c_T^2}d {p} - \frac{\rho_0 \Gamma C_v}{c_T^2}d T,$$
the system becomes
$$
\begin{cases}
-\frac{1}{c_T^2}\frac{\partial^2  p}{\partial t^2} + \frac{\mu}{c_T^2} \frac{\partial}{\partial t} \frac{\partial^2  p}{\partial x^2} + \frac{\partial^2 {p}}{\partial x^2}  = - \frac{\rho_0 \Gamma C_v}{c_T^2}\frac{\partial^2 T}{\partial t^2}  +\frac{\mu\rho_0 \Gamma C_v}{c_T^2}  \frac{\partial}{\partial t}  \left( \frac{\partial^2 T}{\partial x^2}\right) \\
\frac{\partial T}{\partial t} - \nu  \left(\frac{\partial^2 T}{\partial x^2}\right) +  \left(  \frac{ \Gamma^2 C_v T_0} {c_T^2}\frac{\partial T}{\partial t}\right) = \frac{\Gamma T_0}{\rho_0} \frac{1}{c_T^2}\frac{\partial  p}{\partial t}
\end{cases}
$$

The second equation can be simplified by noting that \eqref{A5}
$$
\frac{ \Gamma^2 C_v T_0} {c_T^2} = \frac{ c^2 - c_T^2} {c_T^2} = (\gamma - 1),
$$
and then
$$
\begin{cases}
-\frac{1}{c_T^2}\frac{\partial^2  p}{\partial t^2} + \frac{\mu}{c_T^2} \frac{\partial}{\partial t} \frac{\partial^2  p}{\partial x^2} + \frac{\partial^2 {p}}{\partial x^2}  = - \frac{\rho_0 \Gamma C_v}{c_T^2}\frac{\partial^2 T}{\partial t^2}  +\frac{\mu\rho_0 \Gamma C_v}{c_T^2}  \frac{\partial}{\partial t}  \left( \frac{\partial^2 T}{\partial x^2}\right) \\
\frac{\gamma}{\nu}\frac{\partial T}{\partial t} -  \left(\frac{\partial^2 T}{\partial x^2}\right) = \frac{\Gamma T_0}{\rho_0 \nu} \frac{1}{c_T^2}\frac{\partial  p}{\partial t}
\end{cases}
$$

By eliminating the second space derivative of $T$ in the first equation, we have
$$
\begin{cases}
-\frac{1}{c_T^2}\frac{\partial^2  p}{\partial t^2} + \frac{\mu}{c_T^2} \frac{\partial}{\partial t} \frac{\partial^2  p}{\partial x^2} + \frac{\partial^2 {p}}{\partial x^2}  =  \frac{\rho_0 \Gamma C_v}{c_T^2} (\frac{\gamma \mu}{\nu} - 1) \frac{\partial^2 T}{\partial t^2}  - \frac{ \Gamma^2 C_v T_0}{c_T^4} \frac{\mu}{\nu}    \left(  \frac{\partial^2  p}{\partial t^2} \right) \\
\frac{\gamma}{\nu}\frac{\partial T}{\partial t} -   \left(\frac{\partial^2 T}{\partial x^2}\right) = \frac{\Gamma T_0}{\rho_0 \nu} \frac{1}{c_T^2}\frac{\partial  p}{\partial t}
\end{cases}
$$
and then
$$
\begin{cases}
\left( \frac{(\gamma - 1)\mu}{\nu} -1    \right)\frac{1}{c_T^2}\frac{\partial^2  p}{\partial t^2} + \frac{\mu}{c_T^2} \frac{\partial}{\partial t} \frac{\partial^2  p}{\partial x^2} + \frac{\partial^2 {p}}{\partial x^2}  =  \frac{\rho_0 \Gamma C_v}{c_T^2} (\frac{\gamma \mu}{\nu} - 1) \frac{\partial^2 T}{\partial t^2}  \\
\frac{\gamma}{\nu}\frac{\partial T}{\partial t} -  \left(\frac{\partial^2 T}{\partial x^2}\right) = \frac{\Gamma T_0}{\rho_0 \nu} \frac{1}{c_T^2}\frac{\partial  p}{\partial t}
\end{cases}
$$

We let $\beta_T$ be the expansion coefficient, we have \eqref{A6}
$$
\beta_T =  \frac{1}{v}\frac{\partial v}{\partial T} \big)_p = - \frac{1}{\rho}\frac{\partial \rho}{\partial T} \big)_p = \frac{\Gamma C_v}{ c_T^2 } = \frac{1}{\Gamma T_0} \frac{c^2 - c_T^2}{ c_T^2 } = \frac{\gamma - 1}{\Gamma T_0}
~ \text{, } \alpha = \frac{\nu}{\gamma} \text{  et } c^2 = \gamma c_T^2 
$$
and we deduce the following system of equations: 
$$
\begin{cases}
\left( \frac{(\gamma - 1)\mu}{\gamma \alpha} -1    \right)\frac{1}{c_T^2}\frac{\partial^2  p}{\partial t^2} + \frac{\mu}{c_T^2} \frac{\partial}{\partial t} \frac{\partial^2  p}{\partial x^2} + \frac{\partial^2 {p}}{\partial x^2}  =  \rho_0 \beta_T (\frac{\mu}{\alpha} - 1) \frac{\partial^2 T}{\partial t^2}  \\
\frac{1}{\alpha}\frac{\partial T}{\partial t} -  \left(\frac{\partial^2 T}{\partial x^2}\right) = \frac{\gamma - 1}{\alpha \beta_T \rho_0 c^2 }\frac{\partial  p}{\partial t}
\end{cases}
$$

Again, we look for solutions of the form $Y(t,x) = Y(x) \exp{(h \,t)} $ ( We will take $h = i \omega = 2 i \pi n $ ), we obtain :

\begin{equation}
\left\{\begin{array}{l}
\left(1+\frac{j \omega \gamma \varphi}{\rho v_{s}^{2}}\right) \frac{d^{2} p}{d x^{2}} + 
\frac{\gamma \omega^{2}}{v_{s}^{2}} \left( 1  -  \frac{(\gamma-1)\varphi}{\gamma \,\alpha\, \rho } \right) p=\rho \beta_{T} \omega^{2} \tau  (1-  \frac{\varphi}{\alpha\rho} ) \\
\frac{d^{2} \tau}{d x^{2}}-\frac{j \omega}{\alpha} \tau=-\frac{j \omega(\gamma-1)}{\rho \,\alpha\, \beta_{T}\, v_{s}^{2}} p
\end{array}\right.
\end{equation}
where $v_s = c$, and    $\frac{\varphi}{\rho} = \mu_1 + \mu_2 = \mu$.

The previous system is also equivalent to: 
\begin{equation}
    \left\{\begin{array}{l}
\frac{d^{2} p}{d x^{2}}+\frac{\gamma \varepsilon_{2} \omega^{2}}{v_{s}^{2}} p=\rho \beta_{T} \varepsilon_{1} \omega^{2} \tau \\
\frac{d^{2} \tau}{d x^{2}}-\frac{j \omega}{\alpha} \tau=-\frac{j \omega(\gamma-1)}{\rho \alpha \beta_{T} v_{s}^{2}} p
\end{array}\right.\label{LSN31}
\end{equation}
where
$$
\varepsilon_{1} \equiv \frac{1-\frac{\varphi}{\rho \alpha}}{1+\frac{j \omega \gamma \varphi}{\rho v_{s}^{2}}}, \quad \varepsilon_{2} \equiv \frac{1-\frac{\gamma-1}{\gamma} \frac{\varphi}{\rho \alpha}}{1+\frac{j \omega \gamma \varphi}{\rho v_{s}^{2}}}$$

To simplify the expressions, let :
$$
a=\frac{j \omega}{\alpha}, \quad b=-\frac{j \omega(\gamma-1)}{\rho \alpha \beta_{T} v_{s}^{2}}, \quad c=\rho \beta_{T} \varepsilon_{1} \omega^{2}, \quad d=-\frac{\gamma \varepsilon_{2} \omega^{2}}{v_{s}^{2}}$$
So the system (\ref{LSN31}) becomes:
\begin{equation}
    \left\{\begin{array}{l}
\frac{d^{2} p}{d x^{2}}-d p=c \tau \\
\frac{d^{2} \tau}{d x^{2}}-a \tau=b p
\end{array}\right.
\end{equation}
which implies
\begin{equation}\label{Hupoly}
\left\{\begin{array}{l}
\frac{d^{4} p}{d x^{4}}-(a+d) \frac{d^{2} p}{d x^{2}}+(a d-b c) p=0 \\
\frac{d^{4} \tau}{d x^{4}}-(a+d) \frac{d^{2} \tau}{d x^{2}}+(a d-b c) \tau=0
\end{array}\right.
\end{equation}

We obtain the same bi-laplacien equation as Kirchhoff's equation for the non-dimensional temperature.

\subsection{Stokes-Kirchhoff relation by the Hu approach}\label{sec4.2}

We start by analysing the order of the different terms of the polynomial \eqref{Hupoly} in $\epsilon = \frac{h(\mu+\nu)}{c^2}\ll 1$. For the term $a+d$ we find:
\begin{equation}\label{a+b}
a+d = \frac{c^2 + h(\mu + \nu)}{\frac{\nu}{h}(\frac{c^2}{\gamma} + h\mu)} = \frac{h^2}{c^2} \frac{c^2}{h\nu} \frac{(1+\frac{h(\mu+\nu)}{c^2})}{\frac{1}{\gamma} + \frac{h\nu}{c^2}} = \frac{h^2}{c^2} \frac{\gamma}{\epsilon} + \frac{h^2}{c^2}  o(\frac{1}{\epsilon}),    
\end{equation}
and for $ad - bc$ we find :
$$ad - bc = \frac{h^2}{\frac{\nu}{h}(\frac{c^2}{\gamma} + h\mu)} = \frac{h^4}{c^4} \frac{c^2}{h\nu} \frac{1}{1-\frac{h\mu}{c^2}} = \frac{h^4}{c^4}\frac{1}{\epsilon} + \frac{h^4}{c^4}  o(\frac{1}{\epsilon}).$$

We note that the terms  $a+d$ and $ad - bc$ are of order $\frac{1}{\epsilon}$ and this is important to note for the following, when deriving a Taylor expansion for solutions to \eqref{Hupoly}. This will explain the issue in Hu derivation in \cite{Hu}.

Let us calculate the discriminant of the characteristic polynomial
$$D = (a+d)^2 - 4(ad-bc) $$
$$\sqrt{D} = \sqrt{(a+d)^2 - 4(ad-bc)} = (a+d)\sqrt{1 - 4\frac{ad-bc}{(a+d)^2}}$$
Where
\begin{equation}
4\frac{ad - bc}{(a+d)^2} = 4 \frac{h^2}{\frac{\nu}{h}(\frac{c^2}{\gamma} + h\mu)}  \left(\frac{\frac{\nu}{h}(\frac{c^2}{\gamma} + h\mu)}{c^2 + h(\mu + \nu)}\right)^2
=4\frac{h\nu (\frac{c^2}{\gamma} + h\mu)}{ (c^2 + h(\mu+\nu))^2} 
\end{equation}

To simplify the expression of $\sqrt{D}$, Hu made a first order expansion in $\frac{4h\nu}{\gamma c^2} = O(K_T) = O(\epsilon) \ll 1$ of the term: 
\begin{eqnarray}
\left|\frac{4(a d-b c)}{(a+d)^{2}}\right| = \left|4\frac{h \nu  (\frac{c^2}{\gamma} + h\mu)}{(c^2 + h(\mu+\nu))^2}\right| = \left| 4\frac{ \nu (\frac{h}{c^2\gamma} + \frac{h^2\mu}{c^4})}{\left(1 + \frac{h(\mu+\nu)}{c^2} \right)^2} \right| \nonumber \\
= \left| 4\frac{h\nu }{c^2} \frac{ (\frac{1}{\gamma} + \frac{h \mu}{c^2})}{\left(1 + \frac{h(\mu+\nu)}{c^2} \right)^2}   \right|  = O(\epsilon) \ll 1   
\label{disc}
\end{eqnarray}

This term is therefore of order $1$ in $\epsilon = \frac{h(\mu+\nu)}{c^2}$, and contrary to the statement in \cite{Jordan} this assumption is physically correct in the context of continuous fluid flow, and hence is not the source of error in Hu derivation. Hu then uses an expansion for $\sqrt{D}$ as,
$$ \sqrt{D} = \frac{c^2 + h(\mu + \nu)}{\frac{\nu}{h}(\frac{c^2}{\gamma} + h\mu)} \left(1 - 4 \frac{h \nu (\frac{c^2}{\gamma} + h\mu)}{(c^2 + h(\mu+\nu))^2} \right
)^{1/2} \approx \frac{c^2 + h(\mu + \nu)}{\frac{\nu}{h}(\frac{c^2}{\gamma} + h\mu)} - 2 \frac{1}{\frac{\nu}{h}} \frac{h \nu }{(c^2 + h(\mu+\nu))} $$
$$ \sqrt{D} = \frac{c^2 + h(\mu + \nu)}{\frac{\nu}{h}(\frac{c^2}{\gamma} + h\mu)} - 2\frac{h^2}{c^2 + h(\mu+\nu)} + O(\epsilon)$$
$$
=a+d - 2\frac{h^2}{c^2 + h(\mu+\nu)} + O(\epsilon)
$$
Hence, the roots of the polynomial verify that:
$$k^2 = \frac{1}{2}  \left[
(a+d) \pm \left\{ a+d - 2\frac{h^2}{c^2 + h(\mu+\nu)} \right\} \right]+ O(\epsilon)
$$
$$\begin{cases}
k_1^2 = \frac{h^2}{c^2 + h(\mu+\nu)} = \frac{h^2}{c^2(1 + \frac{h(\mu+\nu)}{c^2})} + O(\epsilon)\\
k_2^2 = (a+d) - \frac{h^2}{c^2 + h(\mu+\nu)}  + O(\epsilon)\end{cases} $$

To obtain the attenuation rate, Hu uses the solution $k_1$ that corresponds to sound waves:
$$k_1^{\pm} = \pm\frac{h}{c} (1 - \frac{h(\mu+\nu)}{2c^2}) = \pm\left\{\frac{i \omega}{c}+\frac{\omega^{2}\gamma \alpha}{2 c^{3}}+\frac{\omega^{2} \varphi}{2 \rho c^{3}}\right\}$$
However this development in not complete in order $1$ on $\epsilon$, and only the order $0$ term is complete. To correct this derivation, the development of the discriminant should be pushed to order $2$ then one will obtain the correct and complete order $1$ development for $k_1$ that is exactly the Stokes-Kirchhoff formula.

$$k_1^{\pm} =\pm \sqrt{ \lambda_{1}}=\pm\left\{\frac{h}{c}-\frac{h^{2}}{2 c^{3}}\left[\mu+\nu\left(1-\frac{1}{\gamma}\right)\right]\right\}=\pm\left\{\frac{i \omega}{c}+\frac{\omega^{2}(\gamma-1) \alpha}{2 c^{3}}+\frac{\omega^{2} \varphi}{2 \rho c^{3}}\right\}$$



Indeed, at order $2$ we have
\begin{align*}
 \sqrt{D} &= \frac{c^2 + h(\mu + \nu)}{\frac{\nu}{h}(\frac{c^2}{\gamma} + h\mu)} \left(1 - 2 \frac{h \nu (\frac{c^2}{\gamma} + h\mu)}{(c^2 + h(\mu+\nu))^2} - 2 \frac{h^2 \nu^2 (\frac{c^2}{\gamma} + h\mu)^2}{(c^2 + h(\mu+\nu))^4}\right)\\
&= a+d - 2\frac{h^2}{c^2 + h(\mu+\nu)} - 2 \frac{h^3\nu(\frac{c^2}{\gamma}+h\mu)}{(c^2 + h(\mu+\nu))^3}
+O(\epsilon^3).\end{align*}
And the roots are given by :
\begin{align*}
    k_1^{2} &=  \frac{h^2}{c^2 + h(\mu+\nu)} +  \frac{h^3\nu(\frac{c^2}{\gamma}+h\mu)}{(c^2 + h(\mu+\nu))^3} +O(\epsilon^2) \\
   &\approx \frac{h^2}{c^2}\left(1-\frac{h(\mu+\nu)}{c^2}\right) + \frac{h^3\nu}{\gamma c^4} +O(\epsilon^2).\\
\end{align*}
Hence, 
\begin{align*}
   k_1^{\pm} &= \pm\left\{ \frac{h}{c} - \frac{h^2}{2c^3}\left[\mu + \nu\left(1 - \frac{1}{\gamma}\right)\right]\right\} +O(\epsilon^2).
\end{align*}

\section{Discussion of the solution form for attenuated plane waves.} \label{sec:disc}

In the previous sections, we seek particular wave solutions to the linearized Navier Stokes equation of the form 
$$
W=W_{0} \exp (i\omega t + k x) = W_{0} \exp ( k_R x) \exp \left(i (\omega t + k_I x) \right)  
$$
with a real sound frequency $\omega$  and a complex wavelength $k=k_R + i k_I$. In this case, $k_R$ corresponds to the attenuation, and the dispersive speed of propagation is give by $$
v_s = \frac{\omega}{|k_I|}.
$$

Physically, such a solution corresponds to the propagation of a perturbation that is maintained at $x=0$ and given by $W(0,t) = W_{0} \exp(i \omega t)$.

In some works, such as the classical work of Stokes \cite{Stokes} and such in \cite{SBJMG}, the authors looked for solutions of the form 
\begin{equation}\label{sol2}
W=W_{0} \exp [i (\omega t + k x)] = W_{0} \exp ( - \omega_I t) \exp \left(i (\omega_R t + k x) \right),  
\end{equation}
Hence with a complex frequency $\omega=\omega_R + i \omega_I$ and a real wavelength $k$. The characteristic polynominal to solve in this case (for $\omega$) can be written :
{\begin{equation}\label{polOmegaC}
\omega^3 - \frac{i \,k^2}{\rho} \left(2\mu + \mu' + \frac{\lambda}{C_v}\right) \omega^2 - \left( c^2 + \frac{(2\mu + \mu')\lambda k^2}{\rho^2 C_v}\right) k^2 \omega + \frac{i \lambda k^4 c^2}{\rho \gamma C_v} = 0
\end{equation}}

In the dimensionless form, we can write :  
{\begin{equation}\label{polOmega}
\left(\frac{\omega}{kc}\right)^3 - \frac{i k}{\rho c} \left(2\mu + \mu' + \frac{\lambda}{C_v}\right) \left(\frac{\omega}{k \,c}\right)^2 - \left( 1 + \frac{(2\mu + \mu')\lambda k^2}{\rho^2 C_v c^2}\right) \left(\frac{\omega}{kc}\right) + \frac{i \lambda k}{\rho c \gamma C_v} = 0
\end{equation}}
or equivalently
{\begin{equation}
\left(\frac{\omega}{kc}\right)^3 - i \left(\kappa + \gamma \kappa_T \right) \left(\frac{\omega}{kc}\right)^2 - \left( 1 + \gamma \kappa \kappa_T \right) \left(\frac{\omega}{kc}\right) + i \kappa_T = 0
\end{equation}}
with the two Knudsen numbers $\kappa =\frac{(2\mu + \mu')k }{\rho c} $ and $\kappa_T = \frac{\lambda k }{\rho c  \, C_p}$.
At first order in $\kappa$ and $\kappa_T$ we have the three solutions :
$$
\xi_1 = 1 + i \frac{\kappa + (\gamma-1)\kappa_T}{2}\quad; \quad\xi_2 = -1 + i \frac{\kappa + (\gamma-1)\kappa_T}{2}\quad ;\quad \xi_3 = i\kappa_T .  
$$
In \ref{wave2} we also give the derivation of the second order terms that corresponds to the dispersion rate in tables \ref{tab1} and \ref{tab2}. We compare in these tables the results obtained in term of attenuation rates and dispersive velocities, in the Stokes case ($Pr=0$) and in the heat conductivity case ($Pr\neq 0$), based in our present work and the works in \cite{Stokes} and \cite{SBJMG}.

\begin{table}\label{tab1}
\begin{tabular}{| L{1.5cm} | C{6cm} | C{6cm} |}
\hline & Attenuation &  Dispersive speed \\
\hline Solution \eqref{sol2} & $$ \omega_I = \frac{ k^2 (2\mu+\mu^{\prime})}{2\rho} +   O(k \,c\,  \kappa^3)$$
$$
= k \,c\, \frac{\kappa}{2} +   O(k \,c\,  \kappa^3) $$  as in \cite{Stokes}. 
&
$$v_s =c-\frac{ k^{2} (2\mu+\mu^{\prime})^{2}}{8\rho^2 c} + O(c \, \kappa^3)$$ 
$$
= c \left( 1 - \frac{\kappa^2}{8}\right) + O(c \, \kappa^3)$$ 
as in \cite{Stokes}. \\
\hline Solution \eqref{sol1} & $$k_R = -\frac{\omega^{2}(2\mu+\mu^{\prime})}{2\rho c^{3}}  + O(\frac{\omega}{c}K_n^3) $$
$$
= -\frac{\omega}{c} \left( \frac{K_n}{2} + O(K_n^3) \right)
$$
& $$ v_s = c +  \frac{3\omega^2(2\mu+\mu^{\prime})^2}{8\rho^2 c^3}  + O(c \,K_n^3)$$
$$
=  c \left( 1 +  \frac{3 K_n^2}{8}  + O( \,K_n^3) \right)
$$\\
\hline
\end{tabular}
\caption{Stokes' case $P_r = 0$ of non conductive fluid. We compare the attenuation and dispersion relations obtained for the two form of solutions \eqref{sol2} and \eqref{sol1}.}
\end{table}

\begin{table}\label{tab2}
\begin{tabular}{| R{1.5cm} | C{6.1cm} | C{7.2cm} |}
\hline & Attenuation &  Dispersive  speed  \\
\hline solution \eqref{sol2} & $$\omega_I = \frac{k^2 (2\mu + \mu')}{2\rho} + \frac{k^2 (\gamma-1)\lambda }{2\rho C_p}$$
$$ = k\,c\, \frac{\kappa + (\gamma-1)\kappa_T}{2}$$
&
{
$$ 
\begin{aligned}
v_s &= c \, \left[1+\frac{(\gamma-1)\kappa\kappa_T-(\gamma-1)\kappa_T^2}{2} \right.\\ 
&-\left.\frac{(\kappa+(\gamma-1)\kappa_T)^2}{8} \right]\\
&= c- \frac{1}{8}\frac{(2\mu + \mu')^2k^2 }{\rho^2 c}\\ &-\frac{\lambda(\gamma-1)(2\mu + \mu')k^2 }{4\rho^2 c C_p} \\
&-\frac{(\gamma-1)(\gamma+3)\lambda^2 k^2 }{8\rho^2 c  \, C_p^2}
\end{aligned}
$$
}
\\
\hline solution \eqref{sol1} & 
{\small
$$k_R = -\frac{\omega^{2}}{2 c^{3}}\left[\frac{2\mu+\mu^{\prime}}{\rho}+\frac{\lambda}{\rho C_v}\left(1-\frac{1}{\gamma}\right)\right]$$ 
$$ = -\frac{\omega}{c} \left( \frac{K_n + (\gamma - 1) K_{th}}{2} \right)$$
}
& $$
\begin{aligned}
v_s &= {c} \left(1 
  +
\frac{3\omega^2(\frac{(2\mu+\mu^{\prime})^2}{\rho^2}+\frac{\lambda^2}{\rho^2 C_v^2})}{8c^4}\right.\\
&+
\frac{\omega^2 (2\mu+\mu^{\prime})\lambda}{4\rho^2 C_v c^4} +
\frac{7\omega^2\lambda^2}{8 \rho^2 C_v^2 \gamma^2 c^4} \\
&-\left.
\frac{\omega^2\lambda(\frac{2\mu+\mu^{\prime}}{\rho}+\frac{\lambda}{\rho C_v})}{2\rho C_v \gamma c^4}  \right) \\
&=1
  + 
\frac{3}{8}{K_n}^2
+
\left(\frac{3}{8}\gamma^2 -\frac{7}{8}-\frac{\gamma}{2}\right)K_{th}^2\\
&+\left(\frac{\gamma}{4}-\frac{1}{2}\right)K_n K_{th}  
\end{aligned}
$$ \\
\hline
\end{tabular}
\caption{Heat conductivity case $P_r \neq 0$. We compare the attenuation and dispersion relations obtained for the two form of solutions \eqref{sol2} and \eqref{sol1}.}
\end{table}
\newpage
Finally, we note that although solutions of the form \eqref{sol2} are purely theoretical as they are initialised with a perfect harmonic solution $W(t=0) = exp(ikx)$ in a dissipative medium, they actually allow to compute the evolution of an arbitrary (unmaintained) perturbation $f(t=0,x) = f_0(x)$ through Fourier transform and using the linearity of the phenomenon considered.
$$
f_0 = \int \hat{f}_0 (k) exp(ikx) dk
 \rightarrow
f(t,x) = \int \hat{f}_0 (k)  \exp ( - \omega_I(k) t) \exp \left(i (\omega_R(k) t + k x) \right) dk.
$$

\section{Conclusion}

As pointed out by \cite{Jordan} and other references, the classical theory presented above based on continuum flow modeling is not sufficient to account for sound attenuation in real fluids, as it gives values much lower than those observed experimentally. This is highlighted by the presentation above where we show that the contributions from the classical Stokes-Kirchhoff theory are of the order of Knudsen numbers for the attenuation, and order two for the dispersion. In fact, other phenomena, such as molecular relaxation processes should be taken into account to explain sound attenuation in fluids outside the low frequency case \cite{A.D}.

On other hand, apart from the classical reference of Kirchhoff in \cite{Kirchhoff} (only available in German), a clear and modern presentation of the Stokes Kirchhoff derivation is missing. We have given above such a presentation and we completed it by giving also the dispersion relation implied by this classical theory.

The authors in \cite{Jordan} has re-established the classical formula of Kirchhoff and pointed out the 'non-correctness' of the alternative formula of \cite{Hu}. However the assumption \eqref{disc} that is questioned is not the source of error, as it is correct and corresponds to the smallness of Knudsen number as imposed by the continuum modeling. In subsection \ref{sec4.2} we explained the error in \cite{Hu}, which is due to the Taylor expansion and the order of magnitude of the different terms (in other words one should pay attention that the term \eqref{a+b} is of order $\frac{1}{\epsilon}$ in the derivation by Hu).

Finally in this paper, we pointed out the different forms of solutions for attenuated harmonic plane waves that one may want to consider. In terms of speed of propagation we showed that the two kind of perturbation linear waves propagate in two different ways: A maintained wave at the source $x=0$ travels at speeds above the thermodynamic speed of sound $c$ (corresponding to the zero frequency limit), while the modes of a local vanishing perturbation travels slower than $c$ . 


\appendix
\section{Some thermodynamic relations}\label{sec:thermo}
We use in this paper the following thermodynamic identities, see \cite{thermo} for a complete presentation. 
\begin{equation}
d e=T d s+\frac{p}{\rho^{2}} d \rho
\label{A0}\end{equation}
\begin{equation}
ds = \frac{C_v}{T} dT - \frac{\Gamma C_v}{\rho} d\rho \label{A1}\end{equation}
\begin{equation}
dp = \rho \Gamma C_v  dT + c_T^2 d\rho  \label{A2}  
\end{equation}

\begin{equation}
c^2 = c_T^2 + \Gamma^2 C_v T     \label{A3}
\end{equation}

\begin{equation}
\alpha = \left. \frac{1}{v}\frac{\partial v}{\partial T} \right)_p = -\left. \frac{1}{\rho}\frac{\partial \rho}{\partial T} \right)_p = \frac{\Gamma C_v}{ c_T^2 } = \frac{1}{\Gamma T_0} \frac{c^2 - c_T^2}{ c_T^2 }\label{A4}
\end{equation}
\begin{equation}
\frac{ \Gamma^2 C_v T_0} {c_T^2} = \frac{ c^2 - c_T^2} {c_T^2} = (\gamma - 1)
    \label{A5}
\end{equation}
\begin{equation}
\beta_T = \left. \frac{1}{v}\frac{\partial v}{\partial T} \right)_p = -\left. \frac{1}{\rho}\frac{\partial \rho}{\partial T} \right)_p = \frac{\Gamma C_v}{ c_T^2 } = \frac{1}{\Gamma T_0} \frac{c^2 - c_T^2}{ c_T^2 } = \frac{\gamma - 1}{\Gamma T_0}
    \label{A6}
\end{equation}

\section{Approximate roots to polynomial \eqref{polOmega} } \label{wave2}

The three roots of \eqref{polOmega} satisfy the equations: 
\begin{eqnarray}
&&\xi_1 \,\xi_2 \,\xi_3 = -i\kappa_T\label{eq1}\\
&&\xi_1 \,\xi_2 + \xi_1 \,\xi_3  + \xi_2\, \xi_3 =-( 1 +\gamma\kappa\kappa_T) \label{eq2}\\
&&\xi_1 + \xi_2 + \xi_3 =  i(\kappa + \gamma\kappa_T)\label{eq3}
\end{eqnarray}

if $\kappa = \kappa_T = 0$, we obtain the order zero solutions:   $$\xi_1 = -1, ~~ \xi_2 = 1, ~~\xi_3 = 0 $$.
If we replace  $\xi_1$ and $\xi_2$ in the equation (\ref{eq1}) then we find at order $1$:
$$\xi_3 = i\kappa_T.$$

The equations (\ref{eq2}) and  (\ref{eq3}) become at order $1$ :
$$\begin{cases}
\xi_1\,\xi_2 + i\kappa_T(\xi_1 + \xi_2) = -(1 + \gamma\kappa\,\kappa_T)\\
\xi_1 + \xi_2 = i(\kappa + \gamma \kappa_T) - i\kappa_T
\end{cases}$$

Hence,
$$\xi_1\,\xi_2 = - 1 - (\gamma-1)\kappa\kappa_T + (\gamma-1)\kappa_T^2  $$ 
We assume to following form for the order $1$ expansion  
$$\xi_1 = -1 + i\alpha + x$$ and   $$\xi_2 = 1 + i\alpha -x$$
with $\alpha$ being the first order term and $x$ the second order term in $\kappa + \kappa_T$. Then 
$$\xi_1\,\xi_2 = -((1-x)^2+\alpha^2) = - 1 - (\gamma-1)\kappa\kappa_T + (\gamma-1)\kappa_T^2, $$
so we have:  $$\alpha = \frac{\xi_1 + \xi_2}{2i} = \frac{\kappa + (\gamma-1)\kappa_T}{2}  $$
so
$$(1-x)^2 =  1 + (\gamma-1)\kappa\kappa_T - (\gamma-1)\kappa_T^2  - \frac{(\kappa + (\gamma-1)\kappa_T)^2}{4}$$
then 
$$x=-\frac{(\gamma-1)\kappa\kappa_T-(\gamma-1)\kappa_T^2}{2}+\frac{(\kappa+(\gamma-1)\kappa_T)^2}{8}.$$

\newpage

\end{document}